\input epsf
\documentstyle[preprint,eqsecnum,aps]{revtex}

\count255=\time\divide\count255 by 60 \xdef\hourmin{\number\count255}
  \multiply\count255 by-60\advance\count255 by\time
  \xdef\hourmin{\hourmin:\ifnum\count255<10 0\fi\the\count255}

\begin{document}


\draft
\preprint{\hbox{JLAB-THY-99-23}}

\title{Radiative Weak Annihilation Decays}

\author{Richard F. Lebed}

\address{Jefferson Lab, 12000 Jefferson Avenue, Newport News, VA 23606}

\date{August, 1999}

\maketitle
\tightenlines
\thispagestyle{empty}

\begin{abstract}
	A class of meson decay modes sensitive to only one quark
topology at leading $G_F$ order (annihilation of valence quarks
through a $W$) is described.  No experimental observations, nor even
upper limits, have been reported for these decays.  This work presents
a simpleminded (order-of-magnitude) calculation of their branching
fractions, and compares to results of previous calculations where
available.  Although rare, one of these modes ($D_s^+ \to \rho^+
\gamma$) might already be observable at charm experiments, two others
($D^+ \to K^{*+} \gamma$, $B^+ \to D_s^{*+} \gamma$) should appear at
the $B$ factories, and the rest at hadron colliders.
\end{abstract}




\newpage
\setcounter{page}{1}

\section{Introduction and Motivation} \label{intro}

	As data on charm and beauty decays accumulates at CLEO, BES,
and the Fermilab and LEP experiments, and soon at the $B$ factories
BABAR and BELLE, it becomes possible to study ever more rare decays in
search of interesting and exotic physics.  This work suggests
examining a largely overlooked class, electromagnetic decays of a
charged meson mediated by the weak annihilation of the meson.  Such
decays are indeed rare, with branching ratios (BRs) suppressed by
$\alpha$, not to mention the difficulty of forcing a camel (the whole
meson wavefunction) through the eye of a needle (the pointlike $W$
vertex multiplied by CKM factors), but their BRs need not be so tiny
as one might think.

\begin{figure}
  \begin{centering}
  \def\epsfsize#1#2{1.50#2}
  \hfil\epsfbox{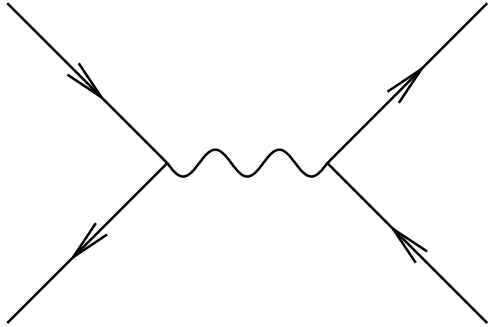}\hfil\hfil

\caption{The $s$-channel weak annihilation topology.\label{fig1}}

  \end{centering}
\end{figure}

\begin{figure}
  \begin{centering}
  \def\epsfsize#1#2{1.50#2}

{\hfil\epsfbox{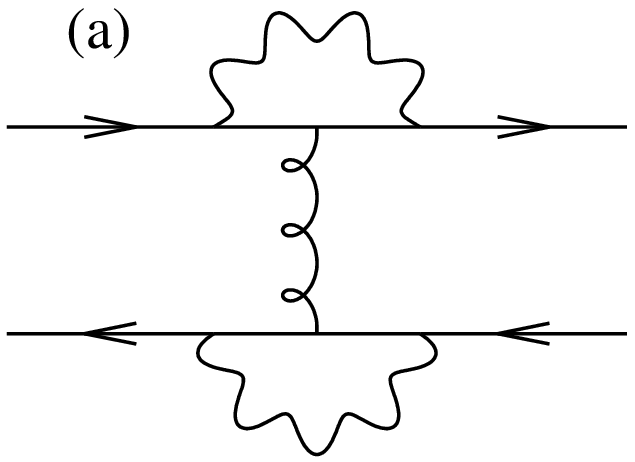}\hfill
\def\epsfsize#1#2{1.50#2}
\epsfbox{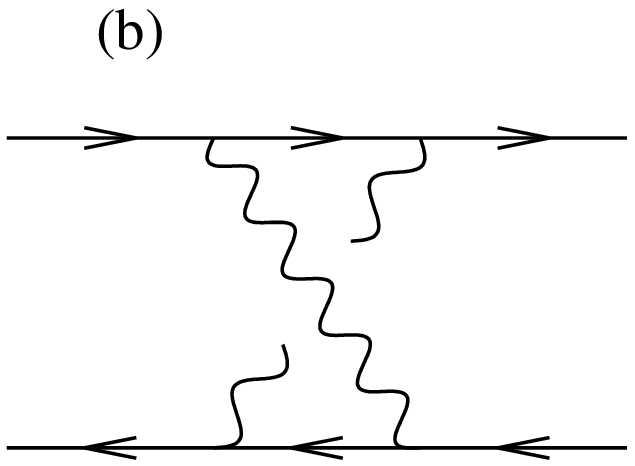}\hfil\hfil}

\caption{$O(G_F^2)$ corrections to Fig.~\ref{fig1}: ($a$) di-penguin
diagram; ($b$) crossed-box diagram.\label{fig2}}

  \end{centering}
\end{figure}

	Let us focus upon processes with only one meson in the final
state.  Such decays are especially interesting because they proceed
through only one weak decay topology at leading order in $G_F$ if $i)$
the flavors of all the valence quarks in both initial and final states
are distinct, and $ii)$ the initial and final quark have the same
electric charge (and similarly for the antiquarks).  This is the
$s$-channel annihilation topology presented in Fig.~\ref{fig1}.  Note
that the first condition requires each valence quark to terminate on a
flavor-changing vertex, while the second eliminates the possibility of
a $t$-channel $W$ exchange.  The photon may be attached to any charged
line, although of course couplings to the lighter constituents are
favored.  At $O(G_F^2)$ corrections enter through diagrams with a
penguin loop on each of the quark and antiquark lines
(Fig.~\ref{fig2}$a$), crossed-box diagrams (Fig.~\ref{fig2}$b$), and
the diagram with the photon coupled to the $W$, none of which is
expected to be very large.  In addition, one may go beyond the valence
diagrams and describe the weak process including its short-distance
QCD corrections in terms of operators mixed through evolution of the
corresponding anomalous dimension matrix,\footnote{Here we refer to
mixing of the usual four-fermi operator with its Fierz reordering.
One finds\cite{BSW,AS} that the coefficient of the original operator
can be enhanced by 20\% or more.} but we do not perform this
refinement in this work.

	The interest in such decays lies partly in the simplicity of
the weak topology and sensitivity to a number of hard-to-isolate CKM
elements (as well as strong and electromagnetic matrix elements), and
partly in the simplicity of the two-body final state.  Indeed, the
hard, monochromatic final-state photon should prove an exceptionally
unambiguous experimental signal for these decays.  It should also be
pointed out that these modes would represent the first electromagnetic
decays observed for the charged $0^-$ mesons $D^+$, $D_s^+$, $B^+$
(except for the famous penguin mode $B^+ \to K^{*+} \gamma$), or
$B_c^+$.  Table~\ref{modes} presents decays representing the 6
possible weak annihilation flavor assignments obeying our constraints,
along with the CKM factors in the amplitude and their behavior in
powers of the Wolfenstein parameter $\lambda \approx 0.22$.

	To our knowledge, the Cabibbo-unsuppressed mode $D_s^+ \to
\rho^+ \gamma$ was first studied in Ref.\cite{AK} via quark model,
then through pole and vector meson dominance (VMD)
calculations\cite{BGHP}, light-cone\cite{KSW}, and effective field
theory techniques\cite{Fajfer}.  Estimates for BR $\times 10^5$ vary
from 2.1\cite{AK} to 80\cite{Fajfer} (see Ref.~\cite{F99} for a
summary).  The present calculation, which for simplicity only takes
into account one pole diagram in the language of
Refs.~\cite{BGHP,Fajfer}, gives $8 \times 10^{-5}$.  The double
Cabibbo-suppressed mode $D^+ \to K^{*+} \gamma$ was also studied in
Refs.~\cite{BGHP,Fajfer}, with BR results ranging from $3$--$30 \times
10^{-7}$; we obtain $6 \times 10^{-7}$.  Encouraged by this
consistency with the more elaborate calculations, we apply our simple
picture to the $B^+$ and $B_c^+$ modes.  While $D_s^+ \to \rho^+
\gamma$ is exciting because it might already appear in charm
experiment data, $D^+ \to K^{*+} \gamma$ is interesting because it
exhibits a neturinoless decay sensitive to $|V_{cs}|$.

	The modes $B^+ \to D^{*+}_s \gamma$ and $D^{*+} \gamma$
(collectively, $D^{*+}_{(s)} \gamma$), were suggested\cite{GL} as
probes of $|V_{ub}|$ and were estimated to have BRs of approximately
$2 \times 10^{-7}$ and $7 \times 10^{-9}$, respectively.  From
Table~\ref{modes} we see that the $B^+$ decays suffer the worst CKM
suppressions.  A number of the theoretical uncertainties associated
with these estimates can be eliminated if the on-shell photon is
replaced by an $\ell^+ \ell^-$ pair\cite{EGN}, and the invariant mass
$q^2$ of the virtual photon is used to define an operator product
expansion.  The price one pays for this improvement is an extra factor
of $\alpha$, so that such decays are estimated to have BRs of a few
times $10^{-10}$ or $10^{-12}$, depending upon the level of Cabibbo
suppression.  The $t$-channel exchange processes mentioned above,
which have neutral initial- and final-state mesons, are discussed in
Ref.~\cite{EGN2}, and yield similar BRs.  Most of these processes are
too rare to be seen in appreciable numbers at the $B$ factories (with
combined yields of some millions of $B^+ B^-$ pairs per
year\cite{Babar}), but may be observable at hadron collider
experiments.

	The $B_c$ modes appear to have been studied only using the
light-cone approach, in Ref.~\cite{AS}.  We obtain BR results several
orders of magnitude larger than theirs, and comment on this
discrepancy below.

	The rest of this paper is organized as follows: In
Sec.~\ref{calc}, we present a very simple-minded calculation of the
rate for these processes and a list of approximations used, while
Sec.~\ref{concl} presents numerical results, outlines experimental
prospects for the observation of these modes, lists potential
theoretical improvements, and concludes.

\section{Calculation} \label{calc}

	The calculation presented in this section is very simplistic,
in that it relies on a number of substantial approximations made
explicit below.  However, it is significant not in providing an exact
determination of widths and BRs, but in obtaining the order of
magnitude of these quantities as an estimate for experimenters
searching for signals of these modes, and as a point of comparison for
theorists performing subsequent, more refined calculations.

\begin{figure}
  \begin{centering}
  \def\epsfsize#1#2{3.00#2}
  \hfil\epsfbox{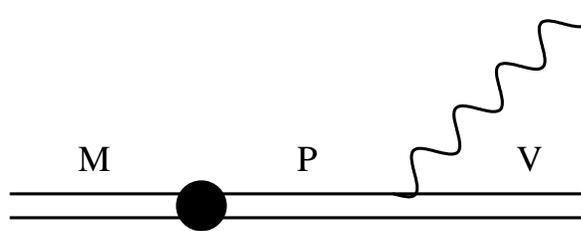}\hfil\hfil

\caption{Meson decay diagram for the process $M \to P \to V\gamma$.
The blob represents the flavor-changing vertex of Fig.~\ref{fig1} and
its corrections.\label{fig3}}

  \end{centering}
\end{figure}

	The generic process we consider is $M \to P \to V \gamma$,
where $M$ is the massive initial $0^-$ state, $P$ is a lighter virtual
$0^-$ meson with the flavor quantum numbers of the final state, and
$V$ is the final-state $1^-$ meson, as depicted in Fig.~\ref{fig3}.
For the charm decays, we have commented that consistency with the more
elaborate calculations in Refs.~\cite{BGHP,Fajfer} indicates that this
elementary ansatz seems to capture the essential order-of-magnitude
physics.  This approach avoids the danger of large cancellations
between competing diagrams, but also runs the risk of missing
important contributions in some cases.  In modeling the decay this
way, we make the following assumptions:

\begin{enumerate}
\item Photon emission from $M$ is neglected, so the process $M \to M^*
\gamma \to V \gamma$ is not included, as was done in
Refs.~\cite{BGHP,Fajfer,GL}.  Indeed, the $D_{(s)}D_{(s)}^*\gamma$
couplings have only measured upper bounds.  In the charm case,
Refs.~\cite{BGHP,Fajfer} used $D_{(s)}D_{(s)}^*\gamma$ input from
previous theoretical calculations.  In Ref.~\cite{GL}, where $M=B^+$,
$M^* = B^{*+}$, $P = D_{(s)}^+$, and $V = D^{*+}_{(s)}$, the
$MM^*\gamma$ and $PV\gamma$ couplings were related through heavy quark
spin-flavor symmetry (HQS) to that of $DD^*\gamma$, and both diagrams
were included.  However, in the current case with, for example, $V =
K^{*+}$ or $\rho^+$, this is no longer an acceptable approximation,
and we include only the photon coupling to the lighter mesons.  This
assumption likely leads to an underestimate of the BR, but probably
not an exceptionally large one: The $K^{*+}$ and $\rho^+$
electromagnetic widths are $50 \pm 5$ and $68 \pm 7$ keV,
respectively, while that of the $D^{*+}$ is less than 4.2
keV;\footnote{Nonetheless, one should note that a small propagator
denominator ($m_V^2 - m_{M^*}^2$) or different a light quark charge in
$M^*$ can enhance the importance of such couplings in the full width.}
moreover, the calculations of Refs.~\cite{BGHP,Fajfer} show the
$D_{(s)}D_{(s)}^*\gamma$ amplitudes to be about 1/4 as large as those
for $PV\gamma$.  Conversely, the light-cone
calculations\cite{AS,KSW} include only the photon coupling to $M$,
arguing that couplings to $V$ are suppressed by light quark masses.

\item Complete factorization with vacuum insertion approximation is
assumed for the weak vertex.  The annihilation of $M$ and the creation
of $P$ are assumed to occur at a single point.  This approximation
neglects both the short-distance QCD corrections as mentioned above,
and long-distance hadronic contributions.  As an example of the
latter, the initial weak vertex may, rather than annihilating the
initial meson, produce a quark (or antiquark) $q$ that is the
antiparticle of one of the meson valence quarks, and this four-quark
intermediate state propagates for some time before $q\bar q$
annihilation occurs.  Specifically, processes like $D^+ \to (K^+ \pi^0
{\rm \ or \ } K^0 \pi^+) \to K^{*+} \gamma$, which require an
understanding of final-state interactions, are not included, nor are
the significant VMD diagrams such as $D^+ \to K^{*+} \rho^0 \to K^{*+}
\gamma$, where the $\rho$ couples resonantly to a photon;
nevertheless, the CKM coefficient of all these diagrams is the same.

\item The intermediate state $P$ is assumed to be the lightest
pseudoscalar with the same flavor quantum numbers as the final-state
$V$.  Certainly many other resonant as well as multiparticle states
with total angular momentum $0$ (that of $M$) can couple the weak
vertex to $V\gamma$; however, the parity-violating couplings to $0^+$
states are neglected here.  The present assumption is made partly
because data exists on the $PV\gamma$ coupling from the observed decay
$V \to P \gamma$, and partly because the lightest pseudoscalar $P$
among all possible intermediates presumably has the one of the largest
couplings to $V\gamma$ due to a relatively large wavefunction overlap.
In any case, this approximation leads to an underestimate of the
correct BR.

\item In comparing the virtual process $P \to V\gamma$ to the on-shell
$V \to P \gamma$, one relates the single (magnetic) form factor ${\cal
C}(q^2)$ at a virtuality of $q^2 = m_M^2 - m_P^2$ to that at $q^2=0$.
We take them numerically equal, although this tends to overestimate
the rate, since form factors tend to fall off away from $q^2=0$.
Nevertheless, we indicate this ratio explicitly in the final
expression, see Eq.~(\ref{rate}) below.

\end{enumerate}

	Given these assumptions, the calculation of the rate is a
simple matter.  The weak mixing vertex, mediated by an operator ${\cal
O}_W$, in vacuum insertion approximation is given by
\begin{eqnarray}
\lefteqn{\left< P(p_M) \left| {\cal O}_W \right| M(p_M) \right>} & &
\nonumber \\ & = & -i \frac{G_F}{\sqrt{2}} V_P V_M B \left< P(p_M)
\left| \overline{Q}_P \gamma^\mu (1-\gamma_5) q^{\vphantom{\dagger}}_P
\right| 0 \right> \left< 0 \left| \bar q^{\vphantom{\dagger}}_M
\gamma_\mu (1-\gamma_5) Q^{\vphantom{\dagger}}_M \right| M(p_M)
\right> \nonumber \\ & = & -i \frac{G_F}{\sqrt{2}} V_P V_M B \left( -i
f^{\vphantom{\dagger}}_P p_M^\mu \right) \left( +i
f^{\vphantom{\dagger}}_M p^{\vphantom{\dagger}}_{M \, \mu} \right)
\nonumber \\ & = & -i \frac{G_F}{\sqrt{2}} V_P V_M
f^{\vphantom{\dagger}}_P f^{\vphantom{\dagger}}_M M^2 B ,
\end{eqnarray}
where $m_M$ is now abbreviated as $M$, the valence structure of $P$ is
$Q^{\vphantom{\dagger}}_P \bar q^{\vphantom{\dagger}}_P$, that of $M$
is $Q^{\vphantom{\dagger}}_M \bar q^{\vphantom{\dagger}}_M$, and
$V_{M,P}$ are the CKM parameters associated with the annihilation of
$M$ and creation of $P$, respectively.  We also allow for a
coefficient $B$ parameterizing the incompleteness of the vacuum
saturation approximation, as in $\bar B B$ mixing, but set it to unity
in our numerical estimates.  The intermediate $0^-$ state $P$ provides
the simple propagator $i/(p_M^2 - m_P^2)$. The photon vertex is
extracted from the decay $\Gamma_{V \to P \gamma}$, which has
invariant amplitude
\begin{equation}
{\cal M} = {\cal C} (p_P^2-m_P^2) \epsilon_{\mu\nu\rho\sigma}
\epsilon^\mu_V \epsilon^{* \nu}_{\gamma} p^\rho_\gamma p^\sigma_V ,
\end{equation}
where the Lorentz coupling is determined in part by the $0^-$ quantum
number of $P$. ${\cal C}$ is a magnetic form factor, and simply
becomes a transition magnetic moment when its momentum transfer
argument $p_P^2-m_P^2$ is set to zero in the on-shell case.  The rate
obtained from this amplitude is
\begin{equation}
\Gamma_{V \to P \gamma} = \frac{1}{12\pi} {\cal C}^2 (0) E_\gamma^3 .
\end{equation}
The full rate for $M \to V \gamma$ also uses ${\cal C}^2$, but now has
the argument $M^2-m_P^2$.  For our numerical estimates, we assume that
${\cal C}$ does not change dramatically over this range, and use data
on $\Gamma_{V \to P \gamma}$ to eliminate ${\cal C}$ from the
expression for $\Gamma_{M \to V \gamma}$; however, since this is
certainly a contentious approximation, we formally retain the ratio of
${\cal C}$ at the two different argument values in the full expression
for the width.  Putting this together, one obtains our central result:
\begin{eqnarray} \label{rate}
\Gamma (M \to V \gamma) & = & \frac 3 2 G_F^2 \left| V_{M} V_{P}
\right|^2 f_M^2 f_P^2 B^2 \, \Gamma_{V \to P \gamma} \left[
\frac{{\cal C} (M^2-m_P^2)}{{\cal C} (0)} \right]^2 \nonumber \\ & &
\times \left( \frac{M^2}{M^2-m_P^2} \right)^2 \left (
\frac{M^2-m_V^2}{m_V^2 - m_P^2} \right)^3 \left(
\frac{m^{\vphantom{\dagger}}_V}{M} \right)^3 .
\end{eqnarray}
Here, the cubed mass factors are nothing more than the ratio of
$E^3_\gamma$ for $M \to V \gamma$ to that for $V \to P \gamma$.

	One may compare Eq.~(\ref{rate}) to Eq.~(17) of Ref.~\cite{GL}
for $B^+ \to D^{*+}_{(s)}$, which in the current notation reads
\begin{eqnarray} \label{hqrate}
\Gamma (M \to V \gamma) & = & \frac{27}{8} G_F^2 \left| V_{M} V_{P}
\right|^2 f_M^4 B^2 \, \Gamma_{V^\prime \to P^\prime \gamma} \left[
\frac{{\cal C} (M^2-m_P^2)}{{\cal C} (0)} \right]^2 \nonumber \\ & &
\times \left[ \frac{m_V M (M-m_V) (M+m_V)^3}{(m_{V^\prime}^2 -
m_{P^\prime}^2)^3} \right] \left(
\frac{m^{\vphantom{\dagger}}_{V^\prime}}{M} \right)^3 ,
\end{eqnarray}
where the $0^-, 1^-$ pair $P^\prime, V^\prime$ are related to $P, V$
by HQS: The primed mesons are introduced when data for $\Gamma_{V \to
P\gamma}$ is unavailable.  In Ref.~\cite{GL}, $P^\prime, V^\prime$
were $D^+, D^{*+}$ and the experimental upper bounds for $D^{*+} \to
D^+ \gamma$ were used; the current calculation uses only information
from the unprimed mesons directly, so $P^\prime, V^\prime \to P, V$
here.  All of the differences between Eqs.~(\ref{rate}) and
(\ref{hqrate}) can be accounted for by the assumptions of HQS: First,
the magnetic moment form factors for all heavy mesons were assumed the
same, except for a trivial coefficient due to the electric charge
$Q_{q(m)}$ of the lighter quark $q$ in the meson $m$ to which the
photon couples.  Since both the $MM^* \gamma$ and $PV\gamma$ couplings
were included in the HQS calculation, an extra enhancement of
$(Q_{q(M)} + Q_{q(V)})^2/Q_{q(V)}^2$, a factor of 9, appeared in
Ref~\cite{GL}.  Next, in HQS one has $f_M^2 M = f_P^2 m_P$, and $m_V -
m_P = O(1/M)$, $M$ being the heavy quark mass.  The remaining
differences arise from the fact that fields containing heavy quarks in
HQS are nearly static, even if the heavy quark changes flavor.  This
leads one to adopt the normalization of HQS states of 1 rather than
$2M$ particles per unit volume, as well as introduce propagators
linear rather than quadratic in particle masses, and these
modifications often lead to effective lowest-order substitutions (in
the current notation) such as $(M+m_V)/2 \to M$.
Equations~(\ref{rate}) and (\ref{hqrate}) are related by the
application of these properties.

\section{Results and Conclusions} \label{concl}

	We employ Eq.~(\ref{rate}) to obtain BR estimates for the 6
modes exhibited in Table~\ref{modes}.  We use standard {\it Review of
Particle Physics\/} (RPP)\cite{pdg} values for masses, decay
constants, CKM elements, and lifetimes whenever possible, with the
following exceptions: We take $|V_{ub}| = 3 \times 10^{-3}$, $f_B =
170$ MeV, and $f_D = f_{D_s} = 200$ MeV.  Following recent
experiments, we use the E791 value\cite{E791} $\tau_{D_s} = 0.518 \pm
0.014 \pm 0.007$ psec, and the CDF values\cite{CDF} $m_{B_c^+} = 6.40
\pm 0.39 \pm 0.13$ GeV, $\tau_{B_c^+} = 0.46^{+0.18}_{-0.16} \pm 0.03$
psec.  OPAL\cite{OPAL} and ALEPH\cite{ALEPH} have also reported a few
$B_c$ candidate events, with mass values consistent with
Ref.~\cite{CDF}.  We also use the most recent lattice
determinations\cite{lat} of $f_{B_c}$, which combined give $425 \pm
11$ MeV.  As mentioned previously, we take $B^2 = 1$ and ${\cal C}^2
(M^2-m_P^2) / {\cal C}^2 (0) = 1$.

	The resulting BRs are exhibited in Table~\ref{res}, along with
the energies of the final-state photon.  We see that the $B^+$ modes
give BRs in agreement with Ref.~\cite{GL}, despite a very different
calculation, and similarly for the $D_{(s)}^+$ decays in comparison
with Refs.~\cite{AK,BGHP,KSW,Fajfer}, as promised in Sec.~\ref{intro}.
A few $D^+ \to K^{*+} \gamma$ and $B^+ \to D_s^{*+} \gamma$ should
appear each year at the $B$ factories, with the exact number depending
upon the correct value of the mantissa in our estimate.  The mode
$D_s^+ \to \rho^+ \gamma$ might even be observable right now at charm
experiments such as at BES (or possibly CLEO) if our estimate is low
by a factor of a few, or the upper bounds of the estimates in
Refs.~\cite{BGHP,Fajfer} are correct, based on limits in the
RPP\cite{pdg}; in any case, existing experiments can place a
meaningful upper limit on its BR.

	However, in contrast to Ref.~\cite{AS}, we find that the
$B_c^+$ modes are rare but not exceptionally so; they find BR$(B_c \to
\rho^+ \gamma) = 8.3 \times 10^{-8}$ and BR$(B_c \to K^{*+} \gamma) =
5.3 \times 10^{-9}$.  This $O(10^3)$ discrepancy might be explainable
if ${\cal C} (0)/ {\cal C} (m_B^2-m_\pi^2) \approx 30$, but then the
order-of-magnitude agreement for $D_{(s)}$ and $B$ decays becomes a
mystery.  Furthermore, even if the form factor ${\cal C}$ falls off
this dramatically, other longer-distance mechanisms (see point 2 in
Sec.~\ref{calc}) would likely step in to maintain the rate.  Of
course, since the $B$ factories are not designed to produce $B_c$'s,
the possible observation of these modes must necessarily wait for the
upcoming hadron collider experiments at the LHC or Tevatron.
 
	Although much physics is neglected in this simple calculation,
our estimates show that weak annihilation decays may be observed in
the near future.  They are attractive from both the experimental and
theoretical perspective.  Improvement of the theoretical calculation
essentially amounts to improving on the four assumptions made in
Sec.~\ref{calc}.  Lifting the first requires new data for the heavy
coupling in $M^* \to M \gamma$, particulary positive data for $D^{*+}
\to D^+ \gamma$.  Improving the vacuum insertion approximation can be
accomplished as in $\bar B B$ mixing, with lattice or model
calculations, and including the short-distance QCD corrections is a
straightforward matter.  Many of the neglected long-distance
corrections such as VMD diagrams have been considered in
Refs.~\cite{BGHP,Fajfer}, but one must take care with their relative
phases, while final-state interactions must still be taken into
account.  As for the remaining two assumptions, one can be freed of
both the lowest-resonance dominance and constant form factor
assumptions by either including other intermediate channels
explicitly, or carrying out light-cone or inclusive quark model
calculations.

	From the experimental side, it would be interesting to see
what direct bounds can be placed on these modes at the current time,
in anticipation of their eventual observation.  Once observed, the
weak annihilation modes will present an interesting probe of CKM
elements, electromagnetic transitions, and meson wavefunctions.

{\samepage
\begin{center}
{\bf Acknowledgments}
\end{center}
I thank C. Carlson, C. Carone, A. Petrov, and V. Sharma for useful
discussions, and G. Burdman and S. Prelov\v{s}ek for pointing out
important references.  This work is supported by the Department of
Energy under contract No.\ DE-AC05-84ER40150.}

\begin{table}

\begin{tabular}{l|cc|cc}

Valence structure & Decay mode && CKM Elements & \\
\hline

$\bar b u \to c \bar s \gamma$ & $B^+ \to D_s^{*+} \gamma$ &&
$V^*_{ub} V^{\vphantom{\dagger}}_{cs} \sim \lambda^3$ & \\

$\bar b u \to c \bar d \gamma$ & $B^+ \to D^{*+} \gamma$ && $V^*_{ub}
V^{\vphantom{\dagger}}_{cd} \sim \lambda^4$ & \\

$\bar b c \to u \bar s \gamma$ & $B_c^+ \to K^{*+} \gamma$ && $V^*_{cb}
V^{\vphantom{\dagger}}_{us} \sim \lambda^3$ & \\

$\bar b c \to d \bar u \gamma$ & $B_c^+ \to \rho^+ \gamma$ && $V^*_{cb}
V^{\vphantom{\dagger}}_{ud} \sim \lambda^2$ & \\

$c \bar d \to u \bar s \gamma$ & $D^+ \to K^{*+} \gamma$ && $V^*_{cd}
V^{\vphantom{\dagger}}_{us} \sim \lambda^2$ & \\

$c \bar s \to u \bar d \gamma$ & $D_s^+ \to \rho^+ \gamma$ && $V^*_{cs}
V^{\vphantom{\dagger}}_{ud} \sim \lambda^0$ &

\end{tabular}

\caption{Flavor structure and mesonic decay modes of weak annihilation
electromagnetic decays.  The CKM coefficient for each process is
accompanied by its magnitude in powers of Wolfenstein $\lambda \approx
0.2$.}
\label{modes}

\end{table}


\begin{table}

\begin{tabular}{l|cc|cc}

Decay mode & BR (est.) && Photon Energy (GeV) & \\
\hline

$B^+ \to D^{*+}_s \gamma$ & $1 \times 10^{-7}$ && 2.22 & \\

$B^+ \to D^{*+} \gamma$ & $7 \times 10^{-9}$ && 2.26 & \\

$B_c^+ \to K^{*+} \gamma$ & $3 \times 10^{-6}$ && 3.14 & \\

$B_c^+ \to \rho^+ \gamma$ & $3 \times 10^{-5}$ && 3.15 & \\

$D^+ \to K^{*+} \gamma$ & $6 \times 10^{-7}$ && 0.72 & \\

$D_s^+ \to \rho^+ \gamma$ & $8 \times 10^{-5}$ && 0.83 &

\end{tabular}

\caption{Estimates of branching ratios for weak annihilation decays
using Eq.~(\ref{rate}).  Also included are energies of the
monochromatic photon.}
\label{res}

\end{table}

\end{document}